# Improvement of superconducting properties by high mixing entropy at blocking layers in BiS$_2$-based superconductor REO$_{0.5}$F$_{0.5}$BiS$_2$

Ryota Sogabe[1], Yosuke Goto[1], Tomohiro Abe[2], Chikako Moriyoshi[2], Yoshihiro Kuroiwa[2], Akira Miura[3], Kiyoharu Tadanaga[3], Yoshikazu Mizuguchi[1]*

*1. Department of Physics, Tokyo Metropolitan University, 1-1, Minami-osawa, Hachioji 192-0397, Japan.*
*2. Department of Physical Science, Hiroshima University, 1-3-1 Kagamiyama, Higashihiroshima, Hiroshima 739-8526, Japan.*
*3. Faculty of Engineering, Hokkaido University, Kita-13, Nishi-8, Kita-ku, Sapporo, Hokkaido 060-8628, Japan.*

\* Corresponding author: Yoshikazu Mizuguchi (mizugu@tmu.ac.jp)

Abstract

To investigate the interlayer interaction in the recently synthesized high-entropy-alloy-type (HEA-type) REO$_{0.5}$F$_{0.5}$BiS$_2$ superconductors (RE: rare earth), we have systematically synthesized two sets of samples with close lattice parameters (close to those of PrO$_{0.5}$F$_{0.5}$BiS$_2$ or CeO$_{0.5}$F$_{0.5}$BiS$_2$) but having different mixing entropy ($\Delta S_{mix}$) for the RE site. The crystal structure was investigated using synchrotron X-ray diffraction and Rietveld refinement. For the examined samples having different $\Delta S_{mix}$, the increase in $\Delta S_{mix}$ does not largely affect the bond lengths and the bond angle of the BiS$_2$ conducting layer but clearly suppresses the in-plane disorder at the in-plane S1 site, which is the parameter essential for the emergence of bulk superconductivity in the REO$_{0.5}$F$_{0.5}$BiS$_2$ system. Bulk nature of superconductivity is improved by the increase in $\Delta S_{mix}$ for the present samples. The results of this work clearly show that the increase in mixing entropy at the blocking layer can positively affect the emergence of bulk superconductivity by modifying the local structure of the conducting layer. This is the evidence of the interaction between the high entropy states of the blocking layers and the physical properties of the conducting layers.

Materials with a layered structure have been extensively studied as a candidate material in which high-temperature superconductivity, unconventional mechanisms of superconductivity, high performance of thermoelectric conversion, and other functional properties [1-20] could emerge. In the material systems like cuprate [1,2], Fe-based [7,8], and BiS$_2$-based [10-18] superconductors, the replacement of a blocking layer, which is alternately stacked with a superconducting (electrically conducting) layer, could increase a transition temperature ($T_c$). Namely, the high flexibility of the layered structure is one of the merits of layered compounds as a functional material. In addition, the flexibility of stacking structure sometimes enables us to modify the crystal structure and superconducting properties. One of the notable examples is the huge pressure effect in the BiS$_2$-based superconductors [11-18]. The dramatic increase in $T_c$ in LaO$_{0.5}$F$_{0.5}$BiS$_2$ suggested the importance of optimization of local structure of the BiS$_2$ layer for the superconductivity in the system [11,17,18].

Recently, we have synthesized BiS$_2$-based superconductors with high-entropy-alloy-type (HEA-type) REO (RE: rare earth) blocking layers [21]. In the HEA-type samples, the RE site of REO$_{0.5}$F$_{0.5}$BiS$_2$ was occupied with five RE elements with a compositional range of 5–35%. The background concept of the study on HEA-type REO$_{0.5}$F$_{0.5}$BiS$_2$ is a normal HEA, which was proposed by Yeh et al. as an alloy containing at least 5 elements with concentrations between 5 and 35 atomic percent [22]. In HEA-type REO$_{0.5}$F$_{0.5}$BiS$_2$, the HEA concept was extended to a layered superconductor system. Notably, the superconducting properties seemed to be improved by making HEA-type REO blocking layer as compared to the BiS$_2$-based superconductors with a conventional (non-HEA) REO blocking layer [21]. However, in the first report on HEA-type REO$_{0.5}$F$_{0.5}$BiS$_2$, the origin of the enhanced bulk characteristics of superconductivity was not clarified. In this study, we investigated the high-entropy (HE) effect to the superconducting properties in REO$_{0.5}$F$_{0.5}$BiS$_2$ from the local structure viewpoint.

The structural concept obtained from this work is summarized in Fig. 1. To discuss the HE effect (the effect of the increase in entropy) generated by mixing RE elements ($\Delta S_{mix}$) at the REO blocking layer to the superconducting states, two sets of REO$_{0.5}$F$_{0.5}$BiS$_2$ samples with similar lattice parameters (similar lattice parameters to those of PrO$_{0.5}$F$_{0.5}$BiS$_2$ or CeO$_{0.5}$F$_{0.5}$BiS$_2$) but having different $\Delta S_{mix}$ were synthesized. $\Delta S_{mix}$ is calculated from the equation of $\Delta S_{mix} = -R \sum_{i=1}^{N} c_i ln c_i$ where $R$, $N$, and $c_i$ are the gas constant, number of the component at the mixed site (RE site here), and the atomic fraction of the component, respectively. As an important condition, carrier doping amount is fixed as 0.5 per Bi by the 50% substitution of O$^{2-}$ by F$^{-}$ at the blocking layer for all the samples. The tuning of lattice parameter was achieved by changing the mixing ration of RE: La, Ce, Pr, Nd, and Sm were used in this study. Notably, as depicted in Fig. 1(b), *in-plane local disorder* of S1 site, whose amplitude is represented as anisotropic displacement parameter $U_{11}$, was clearly suppressed with increasing $\Delta S_{mix}$. According to the suppression of $U_{11}$, the superconducting shielding fraction increased, which corresponds to the improvement of the bulk nature of superconductivity. In our previous works, we have clarified the correlation between the *in-plane chemical pressure* and superconductivity in REO$_{0.5}$F$_{0.5}$BiS$_2$ [23-25]. When the Bi-S1 bond distance decreased, in-plane chemical pressure is enhanced in REO$_{0.5}$F$_{0.5}$BiS$_2$. Furthermore, the in-plane chemical pressure suppresses *in-plane local disorder* in the Bi-S1 plane, which is caused by Bi lone pair [24,25]. By suppressing the in-plane disorder, bulk superconductivity is emerged in REO$_{0.5}$F$_{0.5}$BiS$_2$. However, in the present Pr-based (or Ce-based) REO$_{0.5}$F$_{0.5}$BiS$_2$ samples, in-plane chemical pressure was tuned at almost the same value by tuning the lattice parameter *a* at almost the same. Therefore, to understand the improved superconducting properties by increased $\Delta S_{mix}$ in the present systems, we consider that the HE effect acts like local (chemical) pressure effect. Our results suggest that the HE states in the blocking layer can modify the local structure of the conducting layer and can improve the superconducting properties. Namely, the HE states at the blocking layer can affect the physical properties by modifying the local structure at the electrically conducting layer. This novel concept should be useful for not only $BiS_2$-based superconductors but also developing layered functional materials.

The polycrystalline samples of Pr-based $REO_{0.5}F_{0.5}BiS_2$ with nominal RE = Pr, $Ce_{0.5}Nd_{0.5}$, $Ce_{1/3}Pr_{1/3}Nd_{1/3}$, $La_{0.05}Ce_{0.25}Pr_{0.35}Nd_{0.35}$, and $La_{0.2}Ce_{0.2}Pr_{0.2}Nd_{0.2}Sm_{0.2}$ and Ce-based $REO_{0.5}F_{0.5}BiS_2$ with nominal RE = Ce, $La_{0.4}Pr_{0.6}$, $La_{4/15}Ce_{1/3}Pr_{2/5}$, $La_{0.3}Ce_{0.3}Pr_{0.3}Nd_{0.1}$, $La_{0.34}Ce_{0.34}Pr_{0.2}Nd_{0.06}Sm_{0.06}$ were prepared using the solid-state-reaction method. Powder of $La_2S_3$ (99.9%), $Ce_2S_3$ (99.9%), $Pr_2S_3$ (99.9%), $Nd_2S_3$ (99%), $Sm_2S_3$ (99.9%), $Bi_2O_3$ (99.999%), $BiF_3$ (99.9%), grains of Bi (99.999%), and S (99.99%) were used. Stoichiometric mixtures of the starting chemicals were mixed, pressed into pellets, and heated at 700 °C for 20 h in an evacuated quartz tube. The obtained sample was ground, mixed, pelletized, and heated again under the same heating condition to homogenize the sample. Compositional analysis was performed using energy dispersive X-ray spectroscopy (EDX) with a TM3030 electron microscope system (Hitachi High-Technologies) equipped with a Swift-ED (Oxford) EDX analysis system. The composition analyzed by EDX showed values close to the nominal composition (see the table I). In this paper, for example, the samples of the Pr-based samples were labeled #Pr-1 (RE = Pr), #Pr-2 (RE = $Ce_{0.5}Nd_{0.5}$), #Pr-3 (RE = $Ce_{1/3}Pr_{1/3}Nd_{1/3}$), #Pr-4 (RE = $La_{0.05}Ce_{0.25}Pr_{0.35}Nd_{0.35}$), and #Pr-5 (RE = $La_{0.2}Ce_{0.2}Pr_{0.2}Nd_{0.2}Sm_{0.2}$), according to the number of RE elements contained in the RE blocking layer.

SXRD was performed at the beamline BL02B02, SPring-8 (proposal number: 2018A0074). The wavelength of the X-ray was 0.495274 Å. The SXRD experiments were performed with a sample rotator system at room temperature; the diffraction data were collected using a high-resolution one-dimensional semiconductor detector (multiple MYTHEN system [26]) with a step size of $2\theta = 0.006°$. The crystal structure parameters were refined using the Rietveld method with the RIETAN-FP program [27]. The tetragonal $P4/nmm$ model was used for the Rietveld refinements. The single-phase refinement gave good fitting as shown in Fig. 2, and the obtained structural parameters are summarized in the table I. Crystal structure images were depicted using VESTA software [28].

Figure 3 shows the structural parameters obtained from Rietveld refinements for the #Pr-1, #Pr-2, #Pr-3, #Pr-4, and #Pr-5 samples. Figures 3(a) and 3(b) show the dependences of the lattice parameters $a$ and $c$ on $\Delta S_{mix}$ for the RE site. The lattice parameter $a$ and $c$ are almost independent of $\Delta S_{mix}$ for the #Pr-1, #Pr-2, #Pr-3, #Pr-4, and #Pr-5 samples. In addition, the three Bi-S bond lengths [Fig. 3(c)] and the in-plane S1-Bi-S1 angle [Fig. 3(d)] are also independent of $\Delta S_{mix}$. As mentioned above, the factor essential for the emergence of superconductivity in the $REO_{0.5}F_{0.5}BiS_2$ system is the in-plane chemical pressure [23]. From the refined structural parameters shown in Figs. 3(a-d), the amplitude of the in-plane chemical pressure in the examined samples is almost the same for those samples. Therefore, with those samples, we can investigate the relationship between the local structure, superconductivity, and HE effect (increase in $\Delta S_{mix}$). The displacements for the in-plane Bi and S1 sites were analyzed using anisotropic displacement parameters $U_{11}$ and $U_{33}$ for the in-plane S1 and Bi sites [see the schematic image of Fig. 1(a)]. Figures 3(e) and 3(f) show the $\Delta S_{mix}$ dependences of $U_{11}$ and $U_{33}$ for both the in-plane Bi and S1 sites. The $U_{11}$ for the S1 site is quite large for #Pr-1 as compared to that for Bi and decreases with increasing $\Delta S_{mix}$. The $U_{11}$ for the Bi site does not show a remarkable change with increasing $\Delta S_{mix}$. The decrease in $U_{11}$ of S1 corresponds to the suppression of in-plane disorder at the S1 site, which is essential for the inducement of bulk superconductivity in $REO_{0.5}F_{0.5}BiS_2$ [25]. Namely, the in-plane disorder is suppressed with increasing $\Delta S_{mix}$ while the amplitude of in-plane chemical pressure is almost constant for the system. The $U_{33}$ for S1 increases with increasing $\Delta S_{mix}$. $U_{33}$ for the Bi site shows similar trend to $U_{33}$ for S1.

To investigate the superconducting properties, the temperature dependence of magnetic susceptibility was measured using a superconducting quantum interference devise (SQUID) magnetometer (Quantum design, MPMS-3) with an applied field of 10 Oe after both zero-field cooling (ZFC) and field cooling (FC). Figure 4(a) show the temperature dependences of the magnetic susceptibility $4\pi\chi$ for the Pr-based system. The diamagnetic signals due to the emergence of superconductivity are observed for all the samples. The superconducting transition temperature is almost the same as listed in the table I. The $T_c$ was estimated from the temperature derivative of susceptibility. A superconducting shielding fraction $\Delta 4\pi\chi$ was estimated from the ZFC data at the lowest temperature. $T_c$ does not show a remarkable change as the $\Delta S_{mix}$ increased. This can be explained by the fixed chemical pressure amplitude (*a*-axis value) [23,26]. However, the superconducting shielding fraction $\Delta 4\pi\chi$ increases with increasing $\Delta S_{mix}$, indicating the enhancement of bulk nature of superconductivity by the HE effects [Fig. 4(b)]. The direct correlation between the suppression of $U_{11}$ for the S1 site and $\Delta 4\pi\chi$ is a common trend to the in-plane chemical pressure effect. On the relationship between $U_{33}$ and superconductivity, the presence of positive correlation was observed in a previous work [25]. This may be suggesting the possibility of the superconductivity mechanisms related to the large atomic vibration of in-plane atoms along the *c*-axis, which was recently revealed in a related compounds $LaOBiS_{2-x}Se_x$ [29].

To examine whether the HE effects observed in the Pr-based system is universal for the $REO_{0.5}F_{0.5}BiS_2$ systems with different lattice parameters, we synthesized the Ce-based samples, #Ce-1, #Ce-2, #Ce-3, #Ce-4, and #Ce-5, and investigated the crystal structure and the superconducting properties. As shown in Fig. 5, the samples with similar lattice parameter *a* were obtained. The *c* parameter slightly decreases with increasing $\Delta S_{mix}$, which may be related to the fluctuation of Ce valence [30,31] because the *c* parameter is largely affected by the slight change in carrier concentration. However, the in-plane chemical pressure is basically related to the *a* parameter in this system. As observed in the Pr-based system, $U_{11}$ for the S1 site decreases with increasing $\Delta S_{mix}$ [Fig. 5(e)]. On the superconducting properties, the temperature dependences of magnetic susceptibility $4\pi\chi$ for the #Ce-1, #Ce-2, #Ce-3, #Ce-4, and #Ce-5 samples are plotted in Fig. 6(a). $\Delta 4\pi\chi$ increases with increasing $\Delta S_{mix}$. The $\Delta 4\pi\chi$ for those samples are plotted in Fig. 6(b). Although the data for the #Ce-5 is scattered, other data points indicate the presence of the correlation among $\Delta S_{mix}$, $U_{11}(S1)$, and the superconducting shielding fraction. From the results on the Pr-based and Ce-based systems, we propose that the HE effect can modify the local structure (particularly, in-plane disorder at the S1 site) of the conducting BiS plane of the $REO_{0.5}F_{0.5}BiS_2$ system and can be used for improving superconducting properties of the system.

In conclusion, we have investigated the interaction between the HE states of the REO blocking layer,

the local structure of the electrically conducting $BiS_2$ layer and the superconducting properties of $REO_{0.5}F_{0.5}BiS_2$. The samples with lattice parameters close to $PrO_{0.5}F_{0.5}BiS_2$ (or $CeO_{0.5}F_{0.5}BiS_2$) and largely different mixing entropy ($\Delta S_{mix}$) for the RE site were synthesized by changing the number of RE elements contained in the REO blocking layer. Using synchrotron X-ray diffraction and Rietveld refinements, the crystal structure parameters, including anisotropic displacement parameters, were investigated. The increase in $\Delta S_{mix}$ does not largely affect the bond lengths and the bond angle of the $BiS_2$ conducting layer but clearly suppresses the in-plane disorder at the in-plane S1 site [$U_{11}$ (S1)], which is the parameter essential for the emergence of bulk superconductivity in the $REO_{0.5}F_{0.5}BiS_2$ system. As expected, bulk nature of superconductivity, shielding volume fraction in the superconducting states, is enhanced by the increase in $\Delta S_{mix}$. The present results clearly show that the increase in mixing entropy at the blocking layer can positively affect the emergence of bulk superconductivity in the conducting layer while these layers can be regarded as spatially-separated. The evidence of the interaction between the HE states of the blocking layer, the local structure conducting layer, and the physical properties of the material should be a novel strategy useful to design new layered materials and to enhance the functionality using the concept of high entropy alloy.

Acknowledgements

We thank O. Miura, K. Terashima, and N. L. Saini for their technical supports and fruitful discussion. This study was partially supported by the Grants-in-Aid for Scientific Research (Nos. 15H05886, 16H04493, 16K17944, and 17K19058).

Table I. Structural parameters obtained from Rietveld refinement and superconducting properties for the Pr-based samples (#Pr-1, #Pr-2, #Pr-3, #Pr-4, #Pr-5).

| Label | #Pr-1 | #Pr-2 | #Pr-3 | #Pr-4 | #Pr-5 |
|---|---|---|---|---|---|
| RE (nominal) | Pr | $Ce_{0.5}Nd_{0.5}$ | $Ce_{1/3}Pr_{1/3}Nd_{1/3}$ | $La_{0.05}Ce_{0.25}Pr_{0.35}Nd_{0.35}$ | $La_{0.2}Ce_{0.2}Pr_{0.2}Nd_{0.2}Sm_{0.2}$ |
| RE (EDX) | Pr | $Ce_{0.50}Nd_{0.50}$ | $Ce_{0.34}Pr_{0.34}Nd_{0.32}$ | $La_{0.04}Ce_{0.22}Pr_{0.40}Nd_{0.34}$ | $La_{0.20}Ce_{0.19}Pr_{0.21}Nd_{0.20}Sm_{0.20}$ |
| $\Delta S_{mix}$ (J/Kmol) for RE | 0 | 5.76 | 9.13 | 9.93 | 13.37 |
| Space group | Tetragonal $P4/nmm$ (#129) | | | | |
| $a$ (Å) | 4.01164(5) | 4.01200(5) | 4.01231(6) | 4.01168(4) | 4.01031(4) |
| $c$ (Å) | 13.3683(2) | 13.3632(2) | 13.3629(3) | 13.3679(2) | 13.3847(2) |
| $V$ (Å$^3$) | 215.140(5) | 215.096(5) | 215.124(6) | 215.138(5) | 215.260(4) |
| $R_{wp}$ (%) | 11.4 | 9.3 | 8.9 | 9.4 | 7.8 |
| In-plane Bi-S1 (Å) | 2.83666(4) | 2.83692(5) | 2.83713(5) | 2.83669(3) | 2.83572(3) |
| Interplane Bi-S1 (Å) | 3.323(11) | 3.335(11) | 3.337(14) | 3.341(11) | 3.336(9) |
| Bi-S2 (Å) | 2.516(9) | 2.504(9) | 2.508(9) | 2.505(7) | 2.501(6) |
| S1-Bi-S1 angle (°) | 179.9(5) | 179.7(5) | 179.8(6) | 179.9(5) | 179.9(4) |
| $U_{11}$ (S1) (Å$^2$) | 0.028(4) | 0.019(3) | 0.018(3) | 0.014(3) | 0.010(2) |
| $U_{33}$ (S1) (Å$^2$) | 0.038(7) | 0.062(8) | 0.066(9) | 0.080(8) | 0.060(6) |
| $U_{11}$ (Bi) (Å$^2$) | 0.0110(5) | 0.0111(5) | 0.0100(6) | 0.0111(5) | 0.0103(4) |
| $U_{33}$ (Bi) (Å$^2$) | 0.0238(10) | 0.0295(11) | 0.0307(12) | 0.0321(10) | 0.0243(8) |
| $U$ (S2) (Å$^2$) | 0.010(2) | 0.008(2) | 0.010(2) | 0.012(2) | 0.008(2) |
| $U$ (RE) (Å$^2$) | 0.0121(7) | 0.0093(6) | 0.0091(7) | 0.0108(6) | 0.0096(5) |
| $T_c$ (K) | 3.70 | 3.76 | 3.76 | 3.69 | 3.97 |

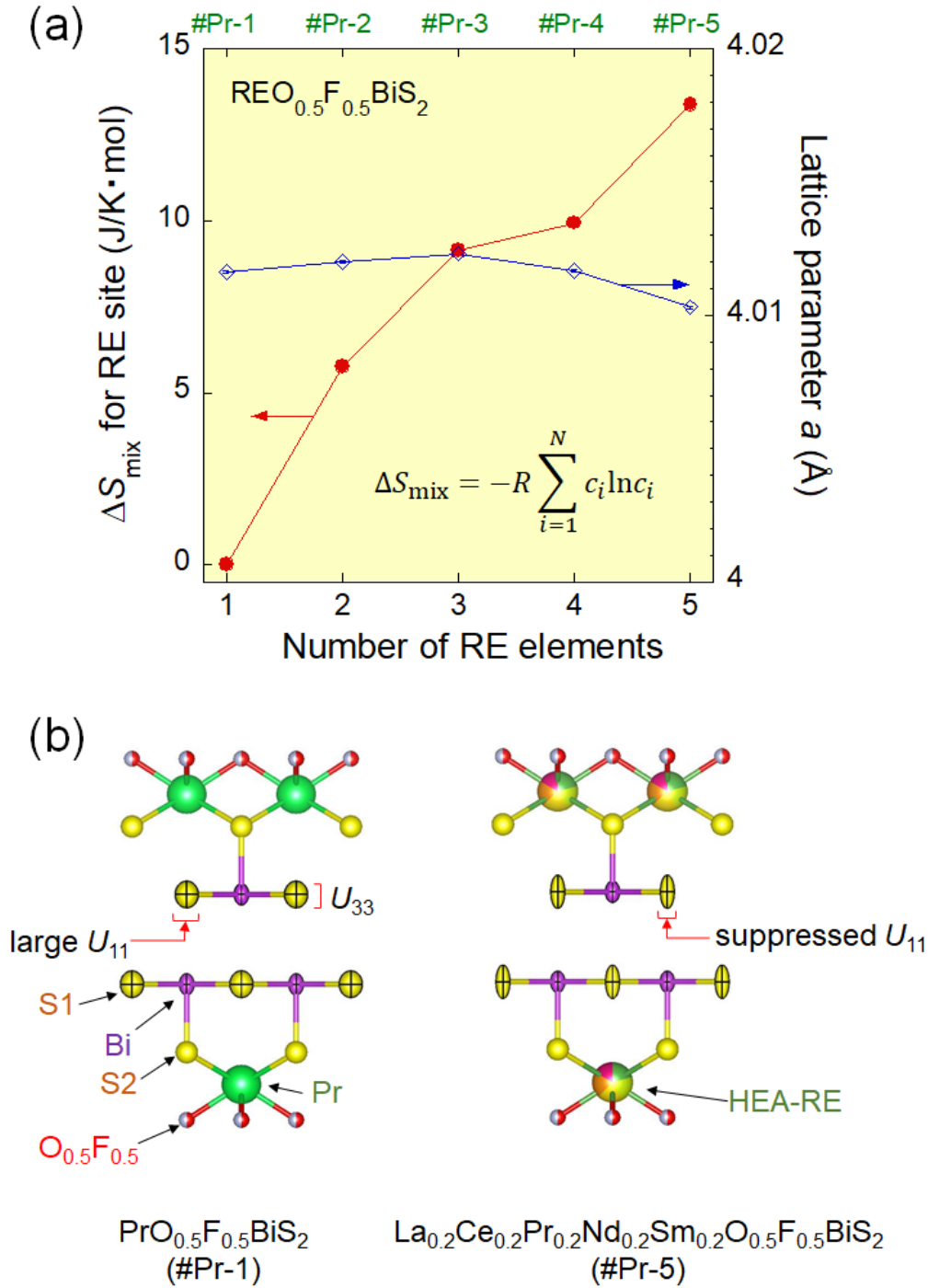

Fig. 1. (Color online) (a) Increased entropy by mixing the RE elements at the REO layer ($\Delta S_{mix}$) and lattice parameter $a$ for $REO_{0.5}F_{0.5}BiS_2$ (#Pr-1, #Pr-2, #Pr-3, #Pr-4, and #Pr-5) whose lattice parameter is close to that of $PrO_{0.5}F_{0.5}BiS_2$ (#Pr-1). (b) Schematic images of the crystal structure for #Pr-1 and #Pr-5 (HEA-type sample) with anisotropic displacement parameters ($U_{11}$ and $U_{33}$) for the in-plane S1 and Bi sites. The thermal ellipsoids for S1 and Bi are depicted with 90% probability. RE and HEA denote rare earth and high entropy alloy, respectively.

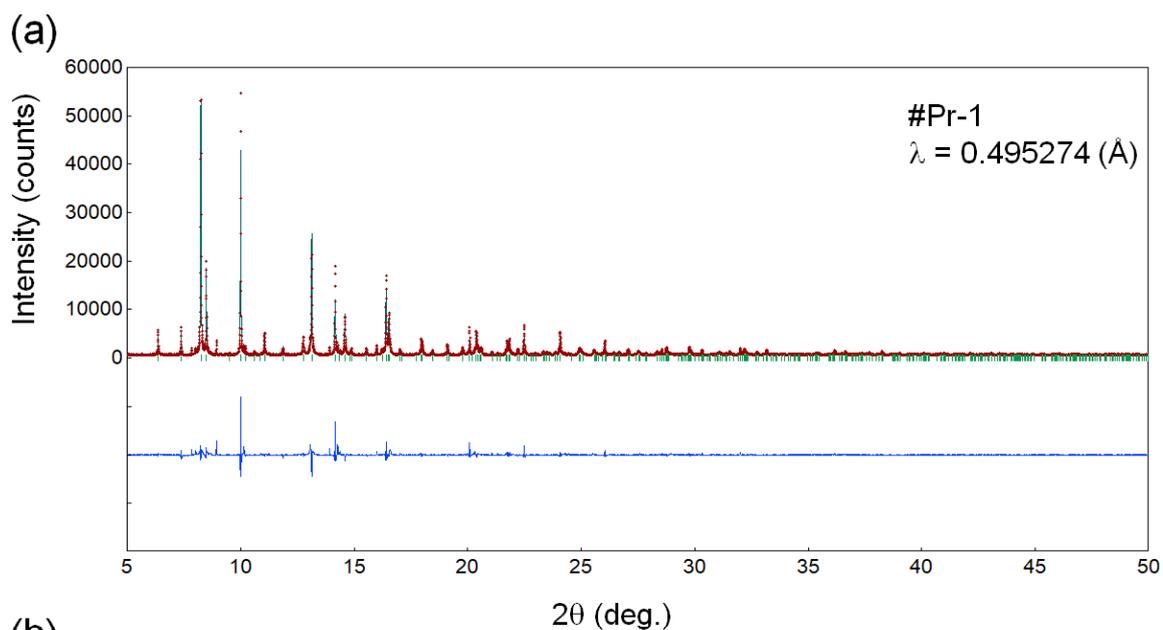
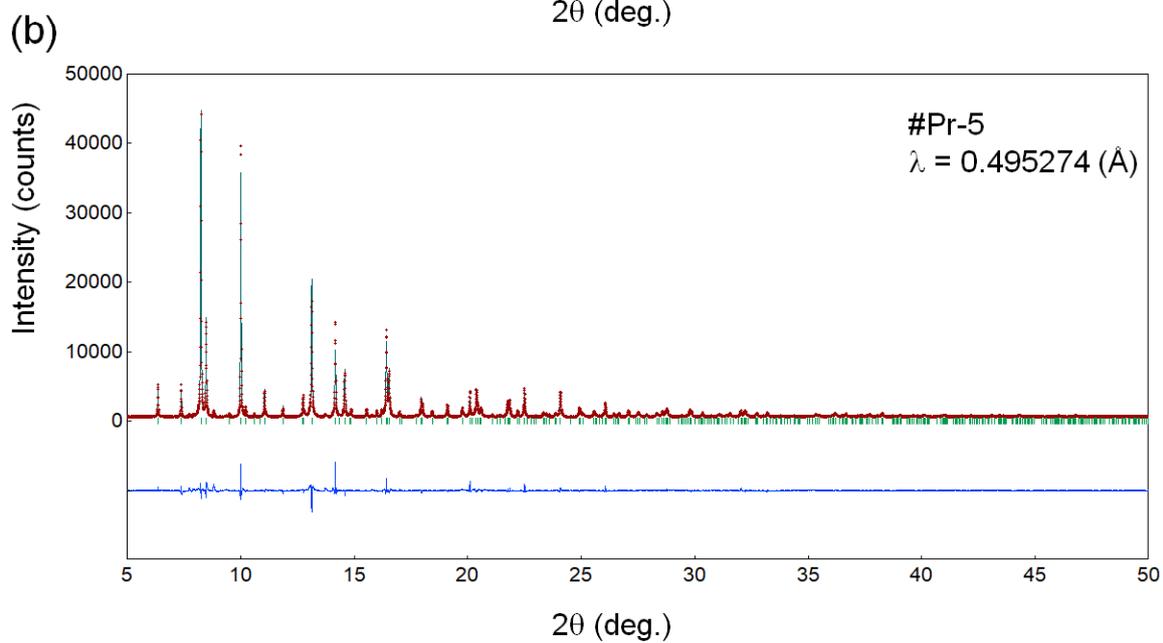

Fig. 2. Synchrotron X-ray diffraction patterns and Rietveld fitting for #Pr-1 and #Pr-5.

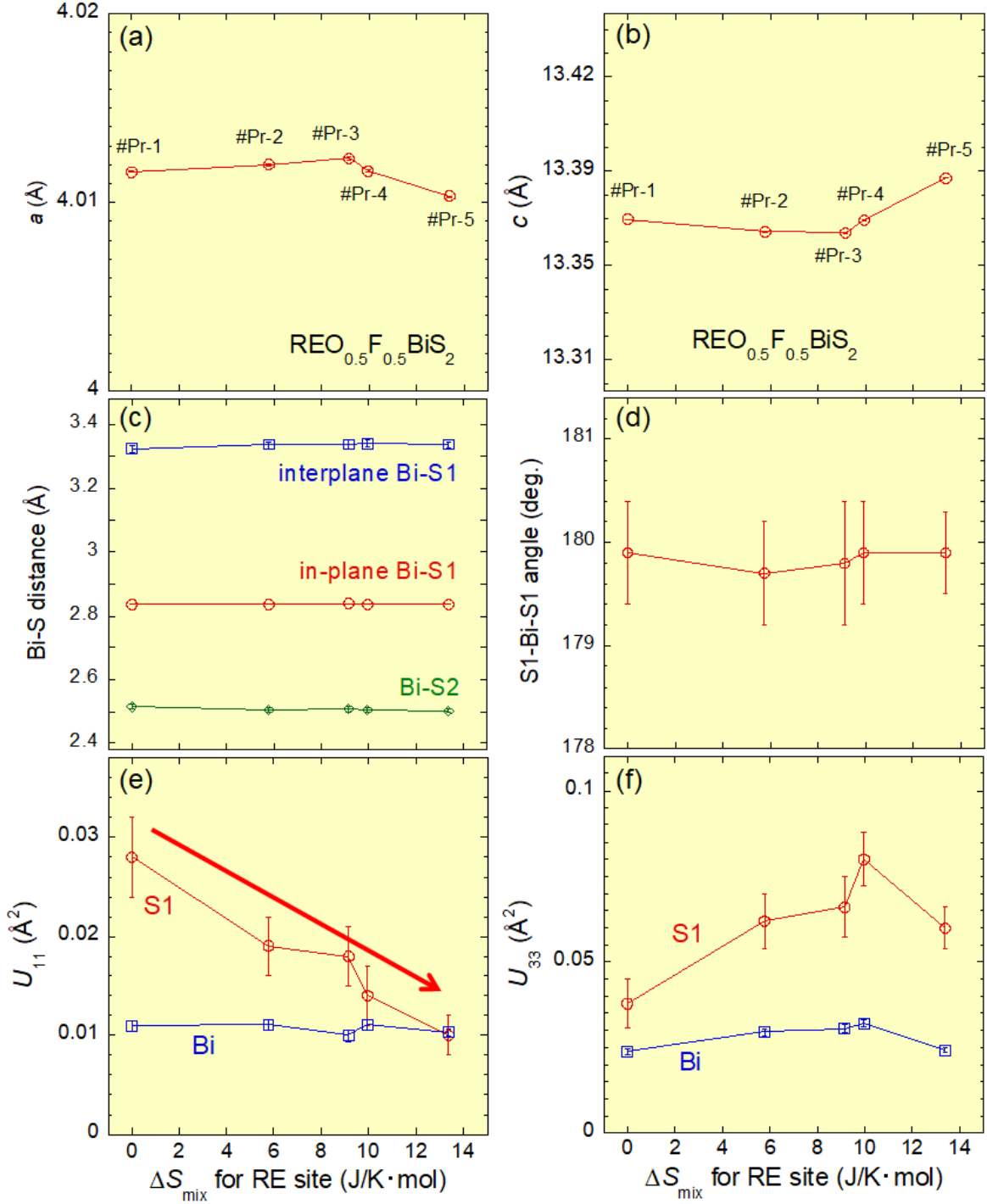

Fig. 3. (Color online) Structural parameters obtained from Rietveld refinement for the #Pr-1, #Pr-2, #Pr-3, #Pr-4, and #Pr-5 samples. $\Delta S_{mix}$ (increase in entropy for RE by mixing RE elements) dependences of (a) lattice parameter $a$, (b) lattice parameter $c$, (c) Bi-S distances, (d) S1-Bi-Si bond angle, (e) anisotropic displacement parameters (in-plane) $U_{11}$, and (f) anisotropic displacement parameters ($c$-axis direction) $U_{33}$ are plotted. The lines in the figures are eye guides.

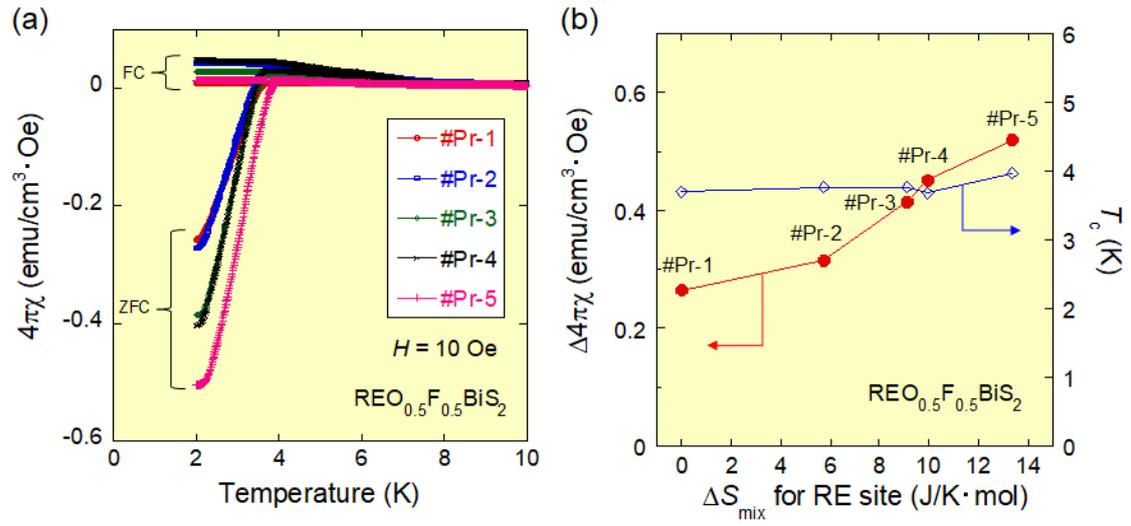

Fig. 4. (Color online) (a) Temperature dependences of magnetic susceptibility $4\pi\chi$ for the #Pr-1, #Pr-2, #Pr-3, #Pr-4, and #Pr-5 samples. (b) $\Delta S_{mix}$ dependences of shielding fraction ($\Delta 4\pi\chi$) and $T_c$ for the #Pr-1, #Pr-2, #Pr-3, #Pr-4, and #Pr-5 samples.

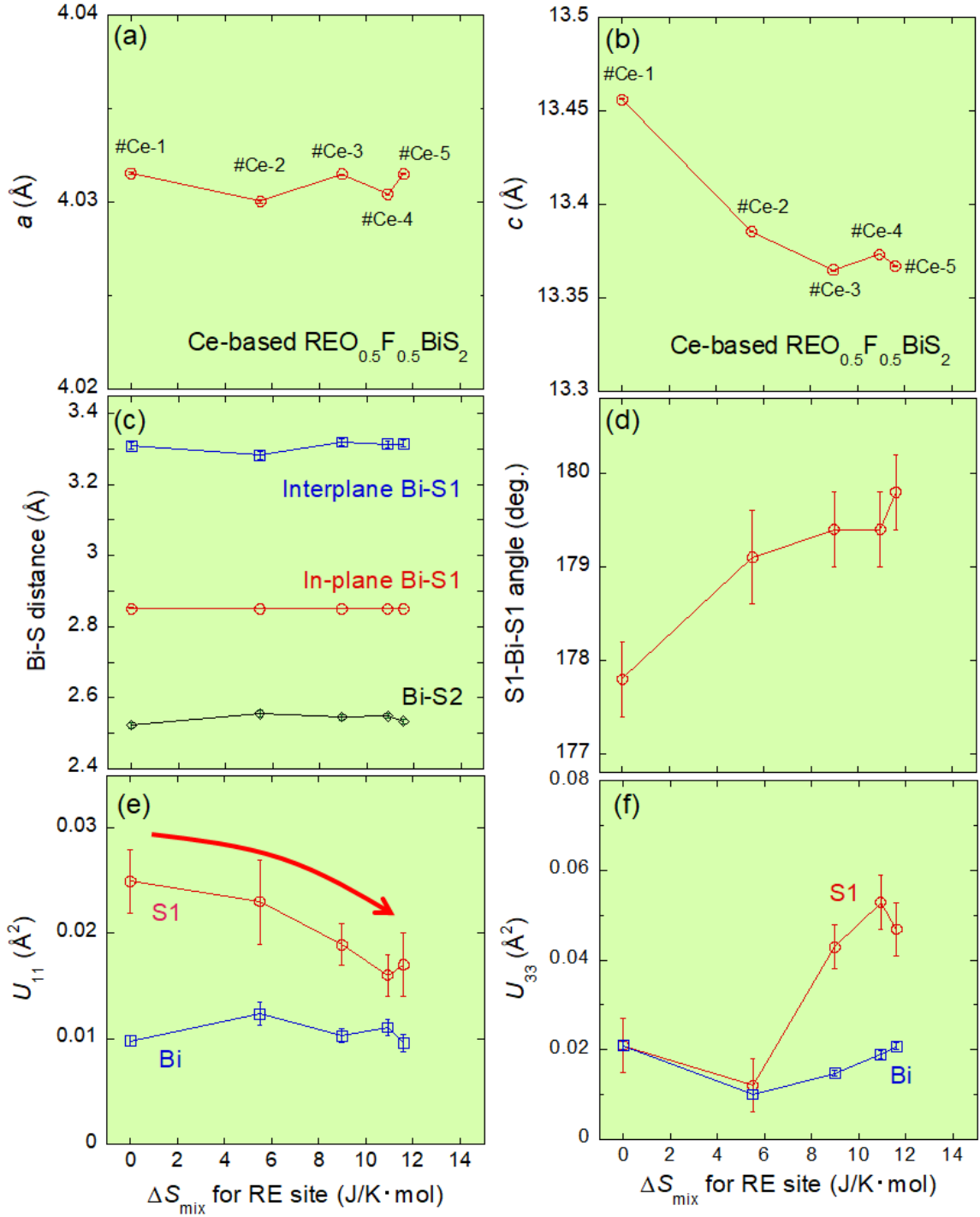

Fig. 5. (Color online) Structural parameters obtained from Rietveld refinement for the #Ce-1, #Ce-2, #Ce-3, #Ce-4, and #Ce-5 samples. $\Delta S_{mix}$ dependences of (a) lattice parameter $a$, (b) lattice parameter $c$, (c) Bi-S distances, (d) S1-Bi-Si bond angle, (e) anisotropic displacement parameters (in-plane) $U_{11}$, and (f) anisotropic displacement parameters ($c$-axis direction) $U_{33}$ are plotted. The lines in the figures are eye guides.

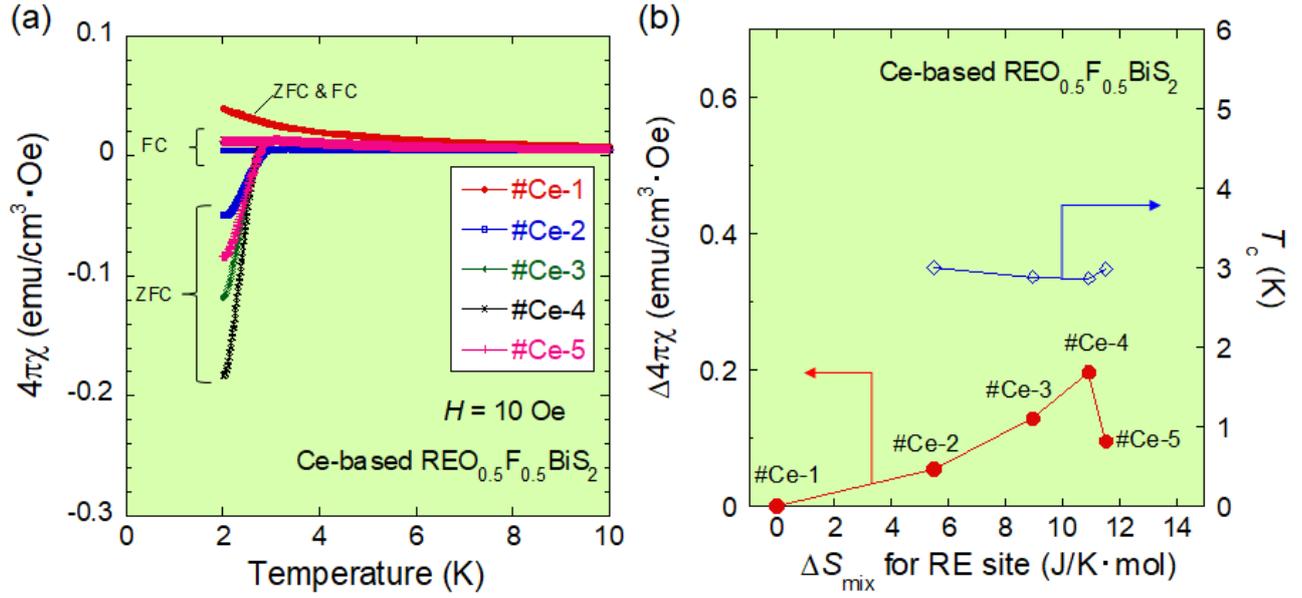

Fig. 6. (Color online) (a) Temperature dependences of magnetic susceptibility $4\pi\chi$ for the #Ce-1, #Ce-2, #Ce-3, #Ce-4, and #Ce-5 samples. (b) $\Delta S_{mix}$ dependences of shielding fraction ($\Delta 4\pi\chi$) and $T_c$ for the #Ce-1, #Ce-2, #Ce-3, #Ce-4, and #Ce-5 samples.